\documentclass[aps,prl,amsmath,amssymb,superscriptaddress,preprintnumbers,twocolumn,
 amsmath,amssymb,
 aps,
]{revtex4-2}

\usepackage{graphicx}% Include figure files
\usepackage{dcolumn}% Align table columns on decimal point
\usepackage{bm}% bold math
\usepackage{cancel}

\usepackage[usenames, dvipsnames]{xcolor}
\usepackage{tikz}
\usetikzlibrary{shapes}
\usepackage[co mpat=1.1.0]{tikz-feynman}
\usepackage{hyperref}% add hypertext capabilities
\newcommand{\SU}{\text{SU}}
\newcommand{\U}{\text{U}}

\begin{document}

%\preprint{.......}

\title{ % The Noble Standard Model via Dualities
 Charting Standard Model Duality and its Signatures
}

\author{Giacomo Cacciapaglia}%
\email{g.cacciapaglia@ip2i.in2p3.fr}
\affiliation{Universite Claude Bernard Lyon 1, CNRS/IN2P3, IP2I UMR 5822, 4 rue Enrico Fermi, F-69100 Villeurbanne, France}
\affiliation{Quantum Theory Center ($\hslash$QTC) at IMADA \& D-IAS, Southern Denmark Univ., Campusvej 55, 5230 Odense M, Denmark}

\author{Francesco Sannino}
\email{sannino@qtc.sdu.dk}
\affiliation{Quantum Theory Center ($\hslash$QTC) at IMADA \& D-IAS, Southern Denmark Univ., Campusvej 55, 5230 Odense M, Denmark}
\affiliation{Dept. of Physics E. Pancini, Universit\`a di Napoli Federico II, via Cintia, 80126 Napoli, Italy}
\affiliation{INFN sezione di Napoli, via Cintia, 80126 Napoli, Italy}
\affiliation{Scuola Superiore Meridionale, Largo S. Marcellino, 10, 80138 Napoli, Italy}

\begin{abstract}
We investigate high and low energy implications of a gauge dual description of the Standard Model.  The high energy electric theory features gauge dynamics involving only fermionic matter fields, while the low energy magnetic description features a quasi-supersymmetric spectrum testable at colliders. The flavour theory is constructed via operators generated at the Planck scale. We further show that duality opens novel avenues for theories of grand unification. 
 \end{abstract}

%\keywords{}%Use showkeys class option if keyword display desired
\maketitle

%\tableofcontents

We show that the Standard Model of particle physics (SM), in its current form, can be interpreted as the magnetic dual of a more fundamental theory. The overall idea of gauge-gauge duality is inspired by Seiberg's work for supersymmetric theories \cite{Seiberg:1994bz,Seiberg:1994pq}, extended to non-supersymmetric cases in \cite{Sannino:2009me,Sannino:2009qc,Mojaza:2011rw}. The ultra-violet (UV) electric description of the Standard Model, at high energy, features only gauge and fermion fields, while the infra-red (IR) magnetic dual contains scalars. We use a non-abelian gauge symmetry of the SM as pivot for the duality, hence intertwining the number of gauge `colours' with the number of matter generations \cite{Sannino:2011mr}.  The SM duality has striking and testable consequences:
\begin{itemize}
    \item The IR magnetic dual features emergent (partial) supersymmetry \cite{Antipin:2011ny}.
    \item Naturally light scalars are required by the duality.
    \item A light gauge-singlet scalar, transforming as a flavour bi-fundamental, can Higgs the electroweak sector and generate flavour structures in the theory.
    \item A universal Yukawa coupling is generated in the magnetic theory.
    \item The scalar-less electric theory spurs new possibilities for grand unified models at high energy.
\end{itemize}
SM dual scenarios are characterised by two or three scales of new physics: a mass scale $\Lambda_S$ for the non-SM composite states in the magnetic theory, expected to appear around few TeVs; the actual duality scale $\Lambda_D$ where the electric theory becomes strong and confines, appearing at intermediate energies between the TeV and Planck scales; if the gauge couplings in the electric theory unify, they would define a third scale $\Lambda_{\rm eGUT}$ below Planck. 

In this letter, 
we present a concrete and experimentally testable scenario of a dual QCD sector within the SM realising all the features listed above. After briefly introducing the duality, we highlight the predictions testable at present and future colliders, as well as model building avenues in flavour physics and grand unification. 

\vspace{0.5cm}

We develop our idea around the duality first presented in \cite{Mojaza:2011rw}, where the UV electric theory is based on a gauged $\SU(N)$ with $N_f$ Dirac fermions in the fundamental representation and one Weyl fermion in the adjoint. Dualities without the adjoint fermion have been discussed in \cite{Sannino:2009me,Sannino:2009qc}, which lead to phenomenological predictions on electroweak observables \cite{Sannino:2010fh}.
Gauge-gauge duality is  mainly based on matching the global anomalies, which depend only on the fermionic matter content. Hence, it is natural to ask whether Seiberg duality applies to non-supersymmetric cousin theories where  elementary scalars in the UV theory are absent but are dynamically generated in the IR. The magnetic and electric theories are expected to describe the same IR physics where the electric gauge coupling becomes strong. The minimal dual theories \cite{Mojaza:2011rw} are illustrated by the field content in Table~\ref{tab:duals}. The numbers of colours in the two theories are related via the flavour symmetry:
\begin{equation}
    X = N_f - N\,.
\end{equation}

\begin{table}[htb]
\centering
\begin{tabular}{|c|c|cccc|}
\hline
\multicolumn{6}{|c|}{Electric theory (UV)} \\
Fields & $\SU(N)$ & $\SU(N_f)_L$ & $\SU(N_f)_R$  & $\U(1)_V$ & $\U(1)_{AF}$\\ \hline\hline
$\lambda$ & Adj & $1$ & $1$  & 0 & $1$ \\
$Q$ & $F$ & $F$ & $1$ & $1$  & $-N/N_f$ \\
$\tilde Q$ & $\bar F$ & $1$ & $\bar F$  & $-1$ & $-N/N_f$ \\ 
%$L$ & $1$ & $F$ & $1$ & $-N$ & 0 \\
%$\tilde{L}$ & $1$ & $1$ & $\bar{F}$ & $N$ & 0 \\
\hline
\hline
\multicolumn{6}{|c|}{Magnetic theory (IR)} \\
Fields & $\SU(X)$ & $\SU(N_f)_L$ & $\SU(N_f)_R$  & $\U(1)_V$ & $\U(1)_{AF}$ \\ \hline\hline
$\lambda_m$ & Adj & $1$ & $1$  & 0 & $1$ \\
$q$ & $F$ & $\bar{F}$ & $1$ & $N/X$  & $-X/N_f$ \\
$\tilde q$ & $\bar F$ & $1$ & $F$  & $-N/X$ & $-X/N_f$  \\
%$l\equiv L$ & $1$ & $F$ & $1$ & $-N$ & $0$\\
%$\tilde{l} \equiv \tilde{L}$ & $1$ & $1$ & $\bar{F}$ & $N$ & $0$ \\
%\hline
$M$ & $1$  & $F$ & $\bar{F}$ & $0$ & $-1+2X/N_f$  \\
\hline
$\phi$ & $F$ & $\bar{F}$ & $1$ & $N/X$  & $1-X/N_f$ \\
$\tilde \phi$ & $\bar F$ & $1$ & $F$  & $-N/X$ & $1-X/N_f$  \\ 
%$\Phi_H$ & $Q  \tilde{Q}$  & $F$ & $\bar{F}$ & $0$ & $-2 +2X/N_f$ \\
$\Phi_H$ & $1$  & $F$ & $\bar{F}$ & $0$ & $2X/N_f$ \\\hline
\end{tabular}
\caption{\label{tab:duals} Matter content for the UV electric (top block) and IR magnetic (bottom block) theories. Only the latter features scalar fields.}
\end{table}

The magnetic theory consists of a magnetic ``gaugino'' $\lambda_m$ in the adjoint representation of $\SU(X)$, $N_f$ Dirac ``quarks'' $q$ and $\tilde{q}$, and a ``mesino'' field $M$. The latter is a gauge singlet, transforming as a bi-fundamental of the flavour symmetry and required by the matching of the global anomalies. Massless coloured scalars $\phi$ and $\tilde{\phi}$ are also required by the decoupling of flavours in the electric theory \cite{Preskill:1981sr}: giving a mass to one flavour of electric quarks is matched in the magnetic dual by a vacuum expectation value (VEV) of the scalars, which breaks $\SU(X) \times \SU(N_f) \to \SU(X-1) \times \SU(N_f-1)$ (i.e., $N_f \to N_f - 1$). Their quantum numbers match those of the electric quarks, where the $\U(1)_{AF}$ charge can be interpreted as the R-charge of the components of supersymmetric multiplets:
\begin{equation}
    (G_\mu, \lambda_m), \qquad (q, \phi), \qquad (\tilde{q}, \tilde{\phi}), \qquad (M, \Phi_H)\,. 
\end{equation}
A light scalar multiplet $\Phi_H$ is, therefore, necessary to accompany the mesinos. Both mesino $M$ and mes-Higgs $\Phi_H$ fields are naturally interpreted as composite states of the electric theory \cite{Sannino:2011mr}:
\begin{equation}
    M = \langle \tilde{Q} \lambda Q \rangle\,, \qquad \Phi_H = \langle \tilde{Q} \lambda \lambda Q\rangle\,.
\end{equation}
Besides kinetic terms and gauge interactions, the magnetic theory allows for the following interactions, invariant under all global symmetries:
\begin{multline} \label{eq:Lmag}
    \mathcal{L}_m \supset y\ q \tilde{q} \Phi_H + y'\ q \tilde{\phi} M + \tilde{y}'\ \tilde{q} \phi M + \\ \xi_L\ \lambda_m q \phi^\dagger + \xi_R\ \lambda_m \tilde{q} \tilde{\phi}^\dagger + \mbox{h.c.}
\end{multline}
and scalar quartic couplings. 
Such couplings are compatible with supersymmetry if the first three emerge from a super-potential coupling ($y=y'=\tilde{y}'$) and the last two from gauge interactions ($\xi_L = \xi_R = g_m$). Note that the magnetic degrees of freedom are weakly interacting at low energies, hence they can naturally be identified as SM states. From the duality point of view, instead, they emerge as deeply composite states of the electric theory when the electric gauge coupling $g_e$ becomes strong at the scale $\Lambda_D$. We therefore pursue the fascinating possibility that one of the asymptotic free gauge groups of the SM, i.e. either QCD's $\SU(3)$ (and its extension to a fourth leptonic colour $\SU(4)$) or the weak $\SU(2)_L$, describes an IR magnetic dual \footnote{Models based on the supersymmetric Seiberg duality have been proposed for QCD's $\SU(3)$ \cite{Maekawa:1995cz,Maekawa:1995ww}, Pati-Salam's $\SU(4)$ \cite{Maekawa:1995nr}, and the weak $\SU(2)$ \cite{Csaki:2011xn,Craig:2011ev} (for AdS/CFT analogues of the latter, see e.g. \cite{Cui:2009dv,Abel:2010vb}).}.

\vspace{0.5cm}

It is tantalising to define the SM duality on QCD gauge symmetry, as the flavour symmetry can be understood as containing the electroweak interactions as follows:
\begin{equation}
    \SU(N_f)_{L/R} \supset \SU(n_g) \otimes SU(2)_{L/R}\,,
\end{equation}
where $n_g = 3$ is the number of generations (leading to $N_f = 6$) and $\SU(2)_R$ is partly   hypercharge gauged. Hence, the mes-Higgs scalar field is composed by bi-doublets of $\SU(2)_L \times \SU(2)_R$, one of which could be identified with the Higgs field. This would lead to a minimal SM duality, described in Table~\ref{tab:QCDdual}. The electric and magnetic gauge groups are the same. As hypercharge is defined as
\begin{equation}
    Y = T^3_R + \frac{1}{6} Q_V\,,
\end{equation}
leptons are introduced as elementary states required by the cancellation of the $U(1)_Y$ gauge anomalies, like in the SM. An alternative possibility, discussed in \cite{Sannino:2011mr}, leads to a more compact duality  by considering leptons as fourth colour \cite{Pati:1974yy}, hence unifying quarks and leptons in $\SU(4)$ multiplets. Here the magnetic dual is a Pati-Salam model which, however, requires to be further broken down to the SM. Models based on supersymmetric Seiberg duality have been proposed for both QCD \cite{Maekawa:1995cz,Maekawa:1995ww} and Pati-Salam \cite{Maekawa:1995nr}. Remarkably  \cite{Sannino:2011mr}, requiring that the electric gauge group remains non abelian, for either $N=3$ or $N=4$, i.e. $N_f - N \geq 2$, sets $n_g=3$ as the smallest number of SM generations compatible with the duality. This lends an elegant dynamical understanding of the need for at least three generations in Nature.

\begin{table}[htb]
\centering
\begin{tabular}{|c|c|cccc|}
\hline
\multicolumn{6}{|c|}{Electric theory (UV)} \\
Fields & $\SU(3)$ & $\SU(6)_L$ & $\SU(6)_R$  & $\U(1)_V$ & $\U(1)_{AF}$\\ \hline\hline
$\lambda$ & Adj & $1$ & $1$  & 0 & $1$ \\
$Q$ & $F$ & $F$ & $1$ & $1$  & $-1/2$ \\
$\tilde Q$ & $\bar F$ & $1$ & $\bar F$  & $-1$ & $-1/2$ \\ 
$L$ & $1$ & $F$ & $1$ & $-3$ & 0 \\
$\tilde{L}$ & $1$ & $1$ & $\bar{F}$ & $3$ & 0 \\
\hline
\hline
\multicolumn{6}{|c|}{Magnetic theory (IR)} \\
Fields & $\SU(3)$ & $\SU(6)_L$ & $\SU(6)_R$  & $\U(1)_V$ & $\U(1)_{AF}$ \\ \hline\hline
$\lambda_m$ & Adj & $1$ & $1$  & 0 & $1$ \\
$q$ & $F$ & $\bar{F}$ & $1$ & $1$  & $-1/2$ \\
$\tilde q$ & $\bar F$ & $1$ & $F$  & $-1$ & $-1/2$  \\
$l\equiv L$ & $1$ & $F$ & $1$ & $-3$ & $0$\\
$\tilde{l} \equiv \tilde{L}$ & $1$ & $1$ & $\bar{F}$ & $3$ & $0$ \\
%\hline
$M$ & $1$  & $F$ & $\bar{F}$ & $0$ & $0$  \\
\hline
$\phi$ & $F$ & $\bar{F}$ & $1$ & $1$  & $1/2$ \\
$\tilde \phi$ & $\bar F$ & $1$ & $F$  & $-1$ & $1/2$  \\ 
$\Phi_H$ & $1$  & $F$ & $\bar{F}$ & $0$ & $1$ \\\hline
\end{tabular}
\caption{\label{tab:QCDdual} Matter content for the UV electric (top block) and IR magnetic (bottom block) dual theories of QCD.}
\end{table}

%\vspace{0.5cm}

The SM magnetic dual in Table~\ref{tab:QCDdual} lists all SM fields plus some new states. In particular, the mes-Higgs field (and similarly the mesino) contains nine complex bi-doublets  \cite{Maekawa:1995cz}:
\begin{equation}
    \Phi_H = \left\{ H_{ij} \right\}\,, \quad i,j = 1,2,3\,,
\end{equation}
with
\begin{equation}
    H_{ij} = (H^u_{ij},\ H^d_{ij})\,,
\end{equation}
where $H^{u,d}$ are distinguished by their hypercharge. In total, the magnetic SM contains 18 ``Higgs'' doublets, one of which should be identified with the SM one. Henceforth, the following states must acquire a mass at around the TeV scale or above: the gaugino $\lambda_m$, the squarks $\phi$ and $\tilde{\phi}$, all mesinos $M$, and all but one of the doublets in $\Phi_H$. The TeV scale is generically required by collider bounds, with the precise values depending on the details of the spectrum. This scale $\Lambda_S$ is the analogue of an emerging supersymmetry breaking scale. The theory is characterised by the following regimes below the Planck scale:
\begin{itemize}
    \item At $E < \Lambda_S$, the SM is valid.
    \item At $\Lambda_s < E < \Lambda_D$, the full spectrum of the magnetic theory appears.
    \item At $E > \Lambda_D$, the electric theory describes the weakly coupled elementary degrees of freedom.
    \item Optionally, at $M_{\rm Pl} > E > \Lambda_{\rm eGUT} > \Lambda_D$, the electric theory can merge into a dual grand unified theory.
\end{itemize}

\begin{figure*}[htb]
\centering
\includegraphics[width=16cm]{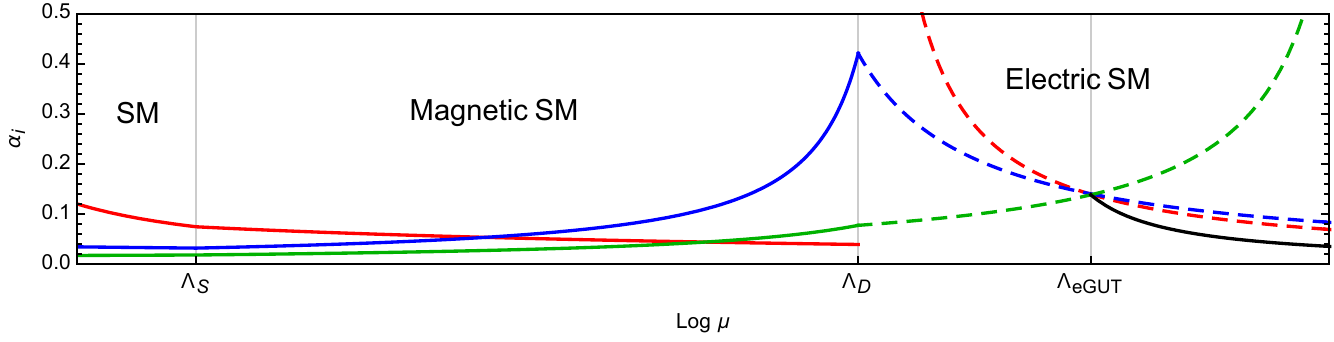}
\caption{\label{fig:running} Schematic evolution of the SM dual gauge couplings from the electroweak to the Planck scales. The three colours refer to $\SU(3)$ (red), $\SU(2)_L$ (blue) and $\U(1)_Y$ (green). Above $\Lambda_{\rm eGUT}$, the three couplings could be replaced by a unified one, here shown in solid black.}
\end{figure*}

\noindent To illustrate this, we show in Fig.~\ref{fig:running} a schematic evolution of the gauge couplings from the electroweak scale up to Planck. At the scale $\Lambda_D$, the electric and magnetic $\SU(3)$ gauge couplings are matched as follows:
\begin{equation}
    \frac{1}{\alpha_e} \sim \alpha_m = \frac{g_m^2}{4\pi}\,.
\end{equation}
The precise estimate of the scale $\Lambda_{\rm eGUT}$  would require higher order studies of the running of the couplings including the impact of the Yukawa interactions Eq.~\eqref{eq:Lmag} in the magnetic regime, which we leave for future studies. Above $\Lambda_D$, only three gauge couplings are present in the electric theory, which simplify enormously the construction of an electric grand unified theory (eGUT).

\vspace{0.5cm}

A crucial aspect of the SM duality, as well as of any model of new physics, is the generation of flavour structures in the quark and lepton masses.
In the magnetic dual, a single Yukawa $y$ is generated, c.f. Eq.~\eqref{eq:Lmag}, while 18 Higgs doublets emerge as composite states within the mes-Higgs field $\Phi_H$. In terms of flavour components, the magnetic Yukawa term can be expanded as
\begin{equation}
    y\ q \tilde{q} \Phi_H = y \sum_{i,j} \left( q_L^i u_R^j H^u_{ij} + q_L^i d_R^j H^d_{ij} \right) \,,
\end{equation}
where $i,j$ are the generation indices.
We stress that this democratic multi-Higgs structure \cite{Hill:2019ldq} is a prediction of the SM duality on QCD. Hierarchical quark masses and mixing can, therefore, be generated by giving hierarchical VEVs to the Higgses, so that the effective Yukawa couplings read
\begin{equation}
    Y^u_{ij} = y \frac{\langle H^u_{ij}\rangle}{v}\quad \mbox{and} \quad  Y^d_{ij} = y \frac{\langle H^d_{ij}\rangle}{v}\,,
\end{equation}
where
\begin{equation}
    v^2 = \sum_{ij} \left(\langle H^u_{ij}\rangle^2 + \langle H^d_{ij}\rangle^2 \right) = \frac{1}{2} (246~\mbox{GeV})^2
\end{equation}
is the SM Higgs VEV. 

Similarly to Ref.~\cite{Hill:2019ldq}, the patterns in the VEVs is generated via suitable mass terms among the 18 Higgs doublets. As an example, the doublet that couples to the top can be identified as the SM Higgs, $H_0 \equiv H^u_{33}$, and it must acquire the largest VEV $\langle H_0 \rangle \approx v$. A smaller VEV can be communicated to the next-to-largest values, i.e. the bottom and charm mass, as well as the mixing between third and second generation. In practice, one adds a mixing of $H^u_{33}$ with a set of `second tier' doublets:
\begin{equation}
    - \mu_b^2 H^u_{33} H^d_{33} - \mu_c^2 H^u_{33} (H^u_{22})^\dagger - \mu_{23}^2 H^u_{33} (H^u_{23} + H^u_{32})^\dagger + \mbox{h.c.} 
\end{equation}
and inducing, therefore, VEVs of order 
\begin{equation}
    \mbox{(2}^{\rm nd}\ \mbox{tier)}\qquad \langle H^d_{33} \rangle = \frac{\mu^2}{M^2_{d,33}} v \sim \frac{m_b}{m_t} v\,,
\end{equation}
and similarly for the others. A third tier defines the strange mass, the second-third generation mixing in the down sector and the second-first mixing in the up sector. The respective Higgses, therefore, mix linearly with the second tier Higgses, and obtain VEVs of the order:
\begin{equation}
    \mbox{(3}^{\rm rd}\ \mbox{tier)}\qquad \langle H \rangle = \frac{\mu^4}{M^4} v\,.
\end{equation}
And so on for the smaller values. Hence, the generation of flavour structures in the Yukawa couplings is translated to the generation of patterns in the mass mixing among the 18 doublets in the magnetic theory.

To simplify the general case, we define a collective notation:
\begin{equation}
    H_\alpha = \left( H^u_{ij},\ (H^d_{ij})^\dagger\right), \;\; \alpha = 0,\dots 17\,,
\end{equation}
which collects 18 doublets with hypercharge $Y_H = 1/2$. We will assume that one Higgs doublet, $H_0 \equiv H^u_{33} $ has an unstable potential and it receives a VEV of order $v$, while the other scalars have (large) positive mass squared and (smaller) linear mixing terms with $H_0$. Hence, the potential reads
\begin{multline}
    V_H = \lambda (H_0^\dagger H_0)^2 - \mu_0^2 (H_0^\dagger H_0) \\ - \mu_{0i}^2 (H_0^\dagger H_i + H_i^\dagger H_0) + M_{ij}^2 (H_i^\dagger H_j)\,,
\end{multline}
where the indices $i,j = 1,\dots 17$ are summed.
We neglect the effect of quartic couplings for the massive doublets, as their VEV will dominantly be induced by the mixing masses $\mu_{0i}^2$. Integrating out the 17 massive states, we obtain:
\begin{equation} \label{eq:masspatt}
    H_i = (M_{ij}^2)^{-1} \mu^2_{0i}\ H_0\,,
\end{equation}
while the $H_0$ potential reproduced the SM one
\begin{equation}
    V_{\rm eff} = \lambda (H_0^\dagger H_0)^2 -  m^2  (H_0^\dagger H_0)\,, 
\end{equation}
with
\begin{equation}
    m^2 = \mu_0^2 + \sum_{i,j=1}^{17} \mu^2_{0j} (M_{ij}^2)^{-1} \mu^2_{0i}\,.
\end{equation}
Flavour hierarchies, therefore, can be generated in terms of matching patterns in the off-diagonal mass terms, $\mu_{0i}^2$ and $\mu^2_{ij} = M^2_{ij}$ for $i\neq j$, and ratios $\frac{\mu}{M} \sim \mathcal{O}(10)$.

In the electric theory, these masses can be written in terms of four-fermion interactions that violate the flavour symmetries. A natural source of such terms is gravity itself, which is expected to break all global symmetries of the theory. Hence, at the Planck scale, we can expect the presence of operators in the form
\begin{equation}
    \mathcal{L}_{\rm Planck} \supset \frac{c^{abcd}}{M_{\rm Pl}^2} (Q^a \tilde{Q}^b)(Q^c \tilde{Q}^d)^\dagger\,.
\end{equation}
At the duality scale $\Lambda_D$, they generate mass terms for the scalars of the order of
\begin{equation}
    \mu^2 \sim \xi \frac{\Lambda_D^4}{M_{\rm Pl}^2}\,,
\end{equation}
where $\xi$ is a numerical coefficient depending on the $c$ couplings and on form factors between the four-fermi interactions and the composite scalars.
Requiring $\mu \sim$~TeV gives an estimate of the `natural' duality scale:
\begin{equation}
    \xi^{-1/4}\ \Lambda_D \sim \sqrt{\mu\,  M_{\rm Pl}} \sim 10^{11}~\mbox{GeV}\,.
\end{equation}

%\vspace{0.5cm}

Lepton masses could be generated via the same mechanism if they are included in the duality via the fourth colour \cite{Sannino:2011mr,Maekawa:1995ww}. In the QCD dual model, they require the presence of another four-fermion interaction in the electric theory:
\begin{equation} \label{eq:LLop}
    \mathcal{L}_e \supset \frac{h_l}{M^2} (L \tilde{L}) (Q \tilde{Q})^\dagger\,,
\end{equation}
which is invariant under the flavour symmetries.
At the duality scale, this will generate a direct coupling to the mes-Higgs field in the form:
\begin{equation}
    \mathcal{L}_m \supset  h_l \mathcal{F}_\Phi \frac{\Lambda_D^2}{M^2} l \tilde{l} \Phi_H^\dagger\,,
\end{equation}
where $\mathcal{F}$ is a form factor.
Once expanded in components, we obtain
\begin{equation}
    l \tilde{l} \Phi_H^\dagger =  \sum_{i,j} \left( l_L^i \nu_R^j (H^d_{ij})^\dagger + l_L^i e_R^j (H^u_{ij})^\dagger \right) \,.
\end{equation}
Interestingly, the largest VEV couples naturally to the $\tau$ lepton, explaining its large mass, while the hierarchy with the top mass is explained by the further suppression given by the scales.
While the operator in Eq.~\eqref{eq:LLop} respects the flavour symmetries in the electric theory, additional flavour structures could be generated via flavour indices in the coupling $h_l$. Finally, large masses for the right-handed neutrinos can be included, close to the scale $\Lambda_D$, to implement a see-saw mechanism of type I.

\vspace{0.5cm}

At current and future colliders, the SM duality can reveal itself via the discovery of new states in the magnetic theory, at the scale $\Lambda_S$. Due to the emerging supersymmetry in the QCD sector, the magnetic gluinos and squarks provide the same signatures as in the minimal supersymmetric SM, hence their masses are constrained to be above roughly 2~TeV at the LHC \cite{ATLAS:2024lda,ATLASSUS,CMSSUS}. A smoking signature is the presence of a large number of higgsino-like states, emerging from the mesino field. If provided a Majorana mass, the lightest mesino could act as dark matter candidate \cite{Belyaev:2022qnf}. Another smoking gun consists of the 17 heavy doublets, responsible for the generation of flavour, however their masses in the multi-TeV range can only be discovered at future higher energy colliders and/or via precision flavour tests. The details of the low energy phenomenology depend on the specific model, and we leave them for future investigation.

\vspace{0.2cm}

In summary, in this letter we investigated the low and high energy implications of a possible dual description of the Standard Model. This provides a new intermediate scale associated with the duality and new directions for theories of grand unification. The emergent supersymmetry at low energies leads to new states, potentially around the TeV scale, which can be produced and studied at current and future colliders. In fact, the non-supersymmetric gauge duality can also be tested via lattice simulation. Finally, the duality investigated here offers a novel origin of the flavour sector of the SM stemming directly from operators generated at the Planck scale.

\section*{Acknowledgements}
We thanks for relevant discussions Aleksandr Azatov, Gia Dvali, Belen Gavela, Tony Gherghetta, Florian Goertz, Tao Han, Manfred Lindner, Luca Vecchi and James Wells. The work of F.S. is
partially supported by the Carlsberg Foundation, semper ardens grant CF22-0922.

\bibliography{biblio}

\begin{thebibliography}{23}%
\makeatletter
\providecommand \@ifxundefined [1]{%
 \@ifx{#1\undefined}
}%
\providecommand \@ifnum [1]{%
 \ifnum #1\expandafter \@firstoftwo
 \else \expandafter \@secondoftwo
 \fi
}%
\providecommand \@ifx [1]{%
 \ifx #1\expandafter \@firstoftwo
 \else \expandafter \@secondoftwo
 \fi
}%
\providecommand \natexlab [1]{#1}%
\providecommand \enquote  [1]{``#1''}%
\providecommand \bibnamefont  [1]{#1}%
\providecommand \bibfnamefont [1]{#1}%
\providecommand \citenamefont [1]{#1}%
\providecommand \href@noop [0]{\@secondoftwo}%
\providecommand \href [0]{\begingroup \@sanitize@url \@href}%
\providecommand \@href[1]{\@@startlink{#1}\@@href}%
\providecommand \@@href[1]{\endgroup#1\@@endlink}%
\providecommand \@sanitize@url [0]{\catcode `\\12\catcode `\$12\catcode `\&12\catcode `\#12\catcode `\^12\catcode `\_12\catcode `\%12\relax}%
\providecommand \@@startlink[1]{}%
\providecommand \@@endlink[0]{}%
\providecommand \url  [0]{\begingroup\@sanitize@url \@url }%
\providecommand \@url [1]{\endgroup\@href {#1}{\urlprefix }}%
\providecommand \urlprefix  [0]{URL }%
\providecommand \Eprint [0]{\href }%
\providecommand \doibase [0]{https://doi.org/}%
\providecommand \selectlanguage [0]{\@gobble}%
\providecommand \bibinfo  [0]{\@secondoftwo}%
\providecommand \bibfield  [0]{\@secondoftwo}%
\providecommand \translation [1]{[#1]}%
\providecommand \BibitemOpen [0]{}%
\providecommand \bibitemStop [0]{}%
\providecommand \bibitemNoStop [0]{.\EOS\space}%
\providecommand \EOS [0]{\spacefactor3000\relax}%
\providecommand \BibitemShut  [1]{\csname bibitem#1\endcsname}%
\let\auto@bib@innerbib\@empty
%</preamble>
\bibitem [{\citenamefont {Seiberg}(1994)}]{Seiberg:1994bz}%
  \BibitemOpen
  \bibfield  {author} {\bibinfo {author} {\bibfnamefont {N.}~\bibnamefont {Seiberg}},\ }\bibfield  {title} {\bibinfo {title} {{Exact results on the space of vacua of four-dimensional SUSY gauge theories}},\ }\href {https://doi.org/10.1103/PhysRevD.49.6857} {\bibfield  {journal} {\bibinfo  {journal} {Phys. Rev. D}\ }\textbf {\bibinfo {volume} {49}},\ \bibinfo {pages} {6857} (\bibinfo {year} {1994})},\ \Eprint {https://arxiv.org/abs/hep-th/9402044} {arXiv:hep-th/9402044} \BibitemShut {NoStop}%
\bibitem [{\citenamefont {Seiberg}(1995)}]{Seiberg:1994pq}%
  \BibitemOpen
  \bibfield  {author} {\bibinfo {author} {\bibfnamefont {N.}~\bibnamefont {Seiberg}},\ }\bibfield  {title} {\bibinfo {title} {{Electric - magnetic duality in supersymmetric nonAbelian gauge theories}},\ }\href {https://doi.org/10.1016/0550-3213(94)00023-8} {\bibfield  {journal} {\bibinfo  {journal} {Nucl. Phys. B}\ }\textbf {\bibinfo {volume} {435}},\ \bibinfo {pages} {129} (\bibinfo {year} {1995})},\ \Eprint {https://arxiv.org/abs/hep-th/9411149} {arXiv:hep-th/9411149} \BibitemShut {NoStop}%
\bibitem [{\citenamefont {Sannino}(2010{\natexlab{a}})}]{Sannino:2009me}%
  \BibitemOpen
  \bibfield  {author} {\bibinfo {author} {\bibfnamefont {F.}~\bibnamefont {Sannino}},\ }\bibfield  {title} {\bibinfo {title} {{Higher Representations Duals}},\ }\href {https://doi.org/10.1016/j.nuclphysb.2009.12.026} {\bibfield  {journal} {\bibinfo  {journal} {Nucl. Phys. B}\ }\textbf {\bibinfo {volume} {830}},\ \bibinfo {pages} {179} (\bibinfo {year} {2010}{\natexlab{a}})},\ \Eprint {https://arxiv.org/abs/0909.4584} {arXiv:0909.4584 [hep-th]} \BibitemShut {NoStop}%
\bibitem [{\citenamefont {Sannino}(2009)}]{Sannino:2009qc}%
  \BibitemOpen
  \bibfield  {author} {\bibinfo {author} {\bibfnamefont {F.}~\bibnamefont {Sannino}},\ }\bibfield  {title} {\bibinfo {title} {{QCD Dual}},\ }\href {https://doi.org/10.1103/PhysRevD.80.065011} {\bibfield  {journal} {\bibinfo  {journal} {Phys. Rev. D}\ }\textbf {\bibinfo {volume} {80}},\ \bibinfo {pages} {065011} (\bibinfo {year} {2009})},\ \Eprint {https://arxiv.org/abs/0907.1364} {arXiv:0907.1364 [hep-th]} \BibitemShut {NoStop}%
\bibitem [{\citenamefont {Mojaza}\ \emph {et~al.}(2011)\citenamefont {Mojaza}, \citenamefont {Nardecchia}, \citenamefont {Pica},\ and\ \citenamefont {Sannino}}]{Mojaza:2011rw}%
  \BibitemOpen
  \bibfield  {author} {\bibinfo {author} {\bibfnamefont {M.}~\bibnamefont {Mojaza}}, \bibinfo {author} {\bibfnamefont {M.}~\bibnamefont {Nardecchia}}, \bibinfo {author} {\bibfnamefont {C.}~\bibnamefont {Pica}},\ and\ \bibinfo {author} {\bibfnamefont {F.}~\bibnamefont {Sannino}},\ }\bibfield  {title} {\bibinfo {title} {{Dual of QCD with One Adjoint Fermion}},\ }\href {https://doi.org/10.1103/PhysRevD.83.065022} {\bibfield  {journal} {\bibinfo  {journal} {Phys. Rev. D}\ }\textbf {\bibinfo {volume} {83}},\ \bibinfo {pages} {065022} (\bibinfo {year} {2011})},\ \Eprint {https://arxiv.org/abs/1101.1522} {arXiv:1101.1522 [hep-th]} \BibitemShut {NoStop}%
\bibitem [{\citenamefont {Sannino}(2011)}]{Sannino:2011mr}%
  \BibitemOpen
  \bibfield  {author} {\bibinfo {author} {\bibfnamefont {F.}~\bibnamefont {Sannino}},\ }\bibfield  {title} {\bibinfo {title} {{The Standard Model is Natural as Magnetic Gauge Theory}},\ }\href {https://doi.org/10.1142/S0217732311036279} {\bibfield  {journal} {\bibinfo  {journal} {Mod. Phys. Lett. A}\ }\textbf {\bibinfo {volume} {26}},\ \bibinfo {pages} {1763} (\bibinfo {year} {2011})},\ \Eprint {https://arxiv.org/abs/1102.5100} {arXiv:1102.5100 [hep-ph]} \BibitemShut {NoStop}%
\bibitem [{\citenamefont {Antipin}\ \emph {et~al.}(2013)\citenamefont {Antipin}, \citenamefont {Mojaza}, \citenamefont {Pica},\ and\ \citenamefont {Sannino}}]{Antipin:2011ny}%
  \BibitemOpen
  \bibfield  {author} {\bibinfo {author} {\bibfnamefont {O.}~\bibnamefont {Antipin}}, \bibinfo {author} {\bibfnamefont {M.}~\bibnamefont {Mojaza}}, \bibinfo {author} {\bibfnamefont {C.}~\bibnamefont {Pica}},\ and\ \bibinfo {author} {\bibfnamefont {F.}~\bibnamefont {Sannino}},\ }\bibfield  {title} {\bibinfo {title} {{Magnetic Fixed Points and Emergent Supersymmetry}},\ }\href {https://doi.org/10.1007/JHEP06(2013)037} {\bibfield  {journal} {\bibinfo  {journal} {JHEP}\ }\textbf {\bibinfo {volume} {06}},\ \bibinfo {pages} {037}},\ \Eprint {https://arxiv.org/abs/1105.1510} {arXiv:1105.1510 [hep-th]} \BibitemShut {NoStop}%
\bibitem [{\citenamefont {Sannino}(2010{\natexlab{b}})}]{Sannino:2010fh}%
  \BibitemOpen
  \bibfield  {author} {\bibinfo {author} {\bibfnamefont {F.}~\bibnamefont {Sannino}},\ }\bibfield  {title} {\bibinfo {title} {{Magnetic S-parameter}},\ }\href {https://doi.org/10.1103/PhysRevLett.105.232002} {\bibfield  {journal} {\bibinfo  {journal} {Phys. Rev. Lett.}\ }\textbf {\bibinfo {volume} {105}},\ \bibinfo {pages} {232002} (\bibinfo {year} {2010}{\natexlab{b}})},\ \Eprint {https://arxiv.org/abs/1007.0254} {arXiv:1007.0254 [hep-ph]} \BibitemShut {NoStop}%
\bibitem [{\citenamefont {Preskill}\ and\ \citenamefont {Weinberg}(1981)}]{Preskill:1981sr}%
  \BibitemOpen
  \bibfield  {author} {\bibinfo {author} {\bibfnamefont {J.}~\bibnamefont {Preskill}}\ and\ \bibinfo {author} {\bibfnamefont {S.}~\bibnamefont {Weinberg}},\ }\bibfield  {title} {\bibinfo {title} {Decoupling constraints on massless composite particles},\ }\href {https://doi.org/10.1103/PhysRevD.24.1059} {\bibfield  {journal} {\bibinfo  {journal} {Phys. Rev. D}\ }\textbf {\bibinfo {volume} {24}},\ \bibinfo {pages} {1059} (\bibinfo {year} {1981})}\BibitemShut {NoStop}%
\bibitem [{Note1()}]{Note1}%
  \BibitemOpen
  \bibinfo {note} {Models based on the supersymmetric Seiberg duality have been proposed for QCD's $\protect \text {SU}(3)$ \cite {Maekawa:1995cz,Maekawa:1995ww}, Pati-Salam's $\protect \text {SU}(4)$ \cite {Maekawa:1995nr}, and the weak $\protect \text {SU}(2)$ \cite {Csaki:2011xn,Craig:2011ev} (for AdS/CFT analogues of the latter, see e.g. \cite {Cui:2009dv,Abel:2010vb}).}\BibitemShut {Stop}%
\bibitem [{\citenamefont {Pati}\ and\ \citenamefont {Salam}(1974)}]{Pati:1974yy}%
  \BibitemOpen
  \bibfield  {author} {\bibinfo {author} {\bibfnamefont {J.~C.}\ \bibnamefont {Pati}}\ and\ \bibinfo {author} {\bibfnamefont {A.}~\bibnamefont {Salam}},\ }\bibfield  {title} {\bibinfo {title} {{Lepton Number as the Fourth Color}},\ }\href {https://doi.org/10.1103/PhysRevD.10.275} {\bibfield  {journal} {\bibinfo  {journal} {Phys. Rev. D}\ }\textbf {\bibinfo {volume} {10}},\ \bibinfo {pages} {275} (\bibinfo {year} {1974})},\ \bibinfo {note} {[Erratum: Phys.Rev.D 11, 703--703 (1975)]}\BibitemShut {NoStop}%
\bibitem [{\citenamefont {Maekawa}(1996)}]{Maekawa:1995cz}%
  \BibitemOpen
  \bibfield  {author} {\bibinfo {author} {\bibfnamefont {N.}~\bibnamefont {Maekawa}},\ }\bibfield  {title} {\bibinfo {title} {{Duality of a supersymmetric standard model}},\ }\href {https://doi.org/10.1143/PTP.95.943} {\bibfield  {journal} {\bibinfo  {journal} {Prog. Theor. Phys.}\ }\textbf {\bibinfo {volume} {95}},\ \bibinfo {pages} {943} (\bibinfo {year} {1996})},\ \Eprint {https://arxiv.org/abs/hep-ph/9509407} {arXiv:hep-ph/9509407} \BibitemShut {NoStop}%
\bibitem [{\citenamefont {Maekawa}\ and\ \citenamefont {Sato}(1996)}]{Maekawa:1995ww}%
  \BibitemOpen
  \bibfield  {author} {\bibinfo {author} {\bibfnamefont {N.}~\bibnamefont {Maekawa}}\ and\ \bibinfo {author} {\bibfnamefont {J.}~\bibnamefont {Sato}},\ }\bibfield  {title} {\bibinfo {title} {{Duality of a supersymmetric standard model without R-parity}},\ }\href {https://doi.org/10.1143/PTP.96.979} {\bibfield  {journal} {\bibinfo  {journal} {Prog. Theor. Phys.}\ }\textbf {\bibinfo {volume} {96}},\ \bibinfo {pages} {979} (\bibinfo {year} {1996})},\ \Eprint {https://arxiv.org/abs/hep-ph/9511395} {arXiv:hep-ph/9511395} \BibitemShut {NoStop}%
\bibitem [{\citenamefont {Maekawa}\ and\ \citenamefont {Takahashi}(1996)}]{Maekawa:1995nr}%
  \BibitemOpen
  \bibfield  {author} {\bibinfo {author} {\bibfnamefont {N.}~\bibnamefont {Maekawa}}\ and\ \bibinfo {author} {\bibfnamefont {T.}~\bibnamefont {Takahashi}},\ }\bibfield  {title} {\bibinfo {title} {{Duality of a supersymmetric model with the Pati-Salam group}},\ }\href {https://doi.org/10.1143/PTP.95.1167} {\bibfield  {journal} {\bibinfo  {journal} {Prog. Theor. Phys.}\ }\textbf {\bibinfo {volume} {95}},\ \bibinfo {pages} {1167} (\bibinfo {year} {1996})},\ \Eprint {https://arxiv.org/abs/hep-ph/9510426} {arXiv:hep-ph/9510426} \BibitemShut {NoStop}%
\bibitem [{\citenamefont {Hill}\ \emph {et~al.}(2019)\citenamefont {Hill}, \citenamefont {Machado}, \citenamefont {Thomsen},\ and\ \citenamefont {Turner}}]{Hill:2019ldq}%
  \BibitemOpen
  \bibfield  {author} {\bibinfo {author} {\bibfnamefont {C.~T.}\ \bibnamefont {Hill}}, \bibinfo {author} {\bibfnamefont {P.~A.~N.}\ \bibnamefont {Machado}}, \bibinfo {author} {\bibfnamefont {A.~E.}\ \bibnamefont {Thomsen}},\ and\ \bibinfo {author} {\bibfnamefont {J.}~\bibnamefont {Turner}},\ }\bibfield  {title} {\bibinfo {title} {{Scalar Democracy}},\ }\href {https://doi.org/10.1103/PhysRevD.100.015015} {\bibfield  {journal} {\bibinfo  {journal} {Phys. Rev. D}\ }\textbf {\bibinfo {volume} {100}},\ \bibinfo {pages} {015015} (\bibinfo {year} {2019})},\ \Eprint {https://arxiv.org/abs/1902.07214} {arXiv:1902.07214 [hep-ph]} \BibitemShut {NoStop}%
\bibitem [{\citenamefont {Aad}\ \emph {et~al.}(2024)\citenamefont {Aad} \emph {et~al.}}]{ATLAS:2024lda}%
  \BibitemOpen
  \bibfield  {author} {\bibinfo {author} {\bibfnamefont {G.}~\bibnamefont {Aad}} \emph {et~al.} (\bibinfo {collaboration} {ATLAS}),\ }\href@noop {} {\bibinfo {title} {{The quest to discover supersymmetry at the ATLAS experiment}}} (\bibinfo {year} {2024}),\ \Eprint {https://arxiv.org/abs/2403.02455} {arXiv:2403.02455 [hep-ex]} \BibitemShut {NoStop}%
\bibitem [{ATL()}]{ATLASSUS}%
  \BibitemOpen
  \href@noop {} {\bibinfo {title} {{ATLAS Supersymmetry searches}}},\ \bibinfo {howpublished} {\url{https://twiki.cern.ch/twiki/bin/view/AtlasPublic/SupersymmetryPublicResults}}\BibitemShut {NoStop}%
\bibitem [{CMS()}]{CMSSUS}%
  \BibitemOpen
  \href@noop {} {\bibinfo {title} {{CMS SUS Physics Results}}},\ \bibinfo {howpublished} {\url{https://twiki.cern.ch/twiki/bin/view/CMSPublic/PhysicsResultsSUS}}\BibitemShut {NoStop}%
\bibitem [{\citenamefont {Belyaev}\ \emph {et~al.}(2022)\citenamefont {Belyaev}, \citenamefont {Cacciapaglia}, \citenamefont {Locke},\ and\ \citenamefont {Pukhov}}]{Belyaev:2022qnf}%
  \BibitemOpen
  \bibfield  {author} {\bibinfo {author} {\bibfnamefont {A.}~\bibnamefont {Belyaev}}, \bibinfo {author} {\bibfnamefont {G.}~\bibnamefont {Cacciapaglia}}, \bibinfo {author} {\bibfnamefont {D.}~\bibnamefont {Locke}},\ and\ \bibinfo {author} {\bibfnamefont {A.}~\bibnamefont {Pukhov}},\ }\bibfield  {title} {\bibinfo {title} {{Minimal consistent Dark Matter models for systematic experimental characterisation: fermion Dark Matter}},\ }\href {https://doi.org/10.1007/JHEP10(2022)014} {\bibfield  {journal} {\bibinfo  {journal} {JHEP}\ }\textbf {\bibinfo {volume} {10}},\ \bibinfo {pages} {014}},\ \Eprint {https://arxiv.org/abs/2203.03660} {arXiv:2203.03660 [hep-ph]} \BibitemShut {NoStop}%
\bibitem [{\citenamefont {Csaki}\ \emph {et~al.}(2011)\citenamefont {Csaki}, \citenamefont {Shirman},\ and\ \citenamefont {Terning}}]{Csaki:2011xn}%
  \BibitemOpen
  \bibfield  {author} {\bibinfo {author} {\bibfnamefont {C.}~\bibnamefont {Csaki}}, \bibinfo {author} {\bibfnamefont {Y.}~\bibnamefont {Shirman}},\ and\ \bibinfo {author} {\bibfnamefont {J.}~\bibnamefont {Terning}},\ }\bibfield  {title} {\bibinfo {title} {{A Seiberg Dual for the MSSM: Partially Composite W and Z}},\ }\href {https://doi.org/10.1103/PhysRevD.84.095011} {\bibfield  {journal} {\bibinfo  {journal} {Phys. Rev. D}\ }\textbf {\bibinfo {volume} {84}},\ \bibinfo {pages} {095011} (\bibinfo {year} {2011})},\ \Eprint {https://arxiv.org/abs/1106.3074} {arXiv:1106.3074 [hep-ph]} \BibitemShut {NoStop}%
\bibitem [{\citenamefont {Craig}\ \emph {et~al.}(2011)\citenamefont {Craig}, \citenamefont {Stolarski},\ and\ \citenamefont {Thaler}}]{Craig:2011ev}%
  \BibitemOpen
  \bibfield  {author} {\bibinfo {author} {\bibfnamefont {N.}~\bibnamefont {Craig}}, \bibinfo {author} {\bibfnamefont {D.}~\bibnamefont {Stolarski}},\ and\ \bibinfo {author} {\bibfnamefont {J.}~\bibnamefont {Thaler}},\ }\bibfield  {title} {\bibinfo {title} {{A Fat Higgs with a Magnetic Personality}},\ }\href {https://doi.org/10.1007/JHEP11(2011)145} {\bibfield  {journal} {\bibinfo  {journal} {JHEP}\ }\textbf {\bibinfo {volume} {11}},\ \bibinfo {pages} {145}},\ \Eprint {https://arxiv.org/abs/1106.2164} {arXiv:1106.2164 [hep-ph]} \BibitemShut {NoStop}%
\bibitem [{\citenamefont {Cui}\ \emph {et~al.}(2009)\citenamefont {Cui}, \citenamefont {Gherghetta},\ and\ \citenamefont {Wells}}]{Cui:2009dv}%
  \BibitemOpen
  \bibfield  {author} {\bibinfo {author} {\bibfnamefont {Y.}~\bibnamefont {Cui}}, \bibinfo {author} {\bibfnamefont {T.}~\bibnamefont {Gherghetta}},\ and\ \bibinfo {author} {\bibfnamefont {J.~D.}\ \bibnamefont {Wells}},\ }\bibfield  {title} {\bibinfo {title} {{Emergent Electroweak Symmetry Breaking with Composite W, Z Bosons}},\ }\href {https://doi.org/10.1088/1126-6708/2009/11/080} {\bibfield  {journal} {\bibinfo  {journal} {JHEP}\ }\textbf {\bibinfo {volume} {11}},\ \bibinfo {pages} {080}},\ \Eprint {https://arxiv.org/abs/0907.0906} {arXiv:0907.0906 [hep-ph]} \BibitemShut {NoStop}%
\bibitem [{\citenamefont {Abel}\ and\ \citenamefont {Gherghetta}(2010)}]{Abel:2010vb}%
  \BibitemOpen
  \bibfield  {author} {\bibinfo {author} {\bibfnamefont {S.}~\bibnamefont {Abel}}\ and\ \bibinfo {author} {\bibfnamefont {T.}~\bibnamefont {Gherghetta}},\ }\bibfield  {title} {\bibinfo {title} {{A slice of AdS$_{5}$ as the large $N$ limit of Seiberg duality}},\ }\href {https://doi.org/10.1007/JHEP12(2010)091} {\bibfield  {journal} {\bibinfo  {journal} {JHEP}\ }\textbf {\bibinfo {volume} {12}},\ \bibinfo {pages} {091}},\ \Eprint {https://arxiv.org/abs/1010.5655} {arXiv:1010.5655 [hep-th]} \BibitemShut {NoStop}%
\end{thebibliography}%

\end{document}